\documentclass[preprint,preprintnumbers,amsmath,amssymb,nofootinbib]{revtex4}
\usepackage{graphicx,color}

\newcommand\nn{\nonumber \\}

\newcommand\br{{\bf r}}
\newcommand\mb{\bar{m}}
\newcommand\nb{\bar{n}}
\newcommand\ntb{\bar{n}_t}

\newcommand{\lk}{\left(}
\newcommand{\rk}{\right)}

\newcommand{\ldk}{\left[}

\newcommand{\rdk}{\right]}
\newcommand\beq{ \begin{eqnarray} }
\newcommand\eeq{ \end{eqnarray} }

\begin{document}

\title{BEC polaron in harmonic trap potentials in the weak coupling regime: 
Lee-Low-Pines-type approach}
\author{Eiji Nakano${}^1$}
\email{e.nakano@kochi-u.ac.jp}
\author{Hiroyuki Yabu${}^2$, and Kei Iida${}^1$}
\affiliation{${}^1$Department of Physics, Kochi University, Kochi 780-8520, Japan}
\affiliation{${}^2$Department of Physics, Ritsumeikan University, Kusatsu, Siga 525-8577, Japan}

\date{\today}

\begin{abstract}
We have calculated the zero-temperature binding energy of a single 
impurity atom immersed in a Bose-Einstein condensate (BEC) of 
ultracold atoms. The impurity and the condensed atoms are 
trapped in the respective axially symmetric harmonic potentials, 
where the impurity interacts with bosonic atoms 
in the condensate via low-energy $s$-wave scattering.  In this case, 
bosons are excited around the impurity to form a quasiparticle, namely, 
a BEC polaron.  We have developed a variational method, 
{\it a la} Lee-Low-Pines (LLP), 
for description of the polaron that has a conserved angular 
momentum around the symmetric axis.  
We find from numerical results 
that the binding between the impurity and the excited bosons breaks  
the degeneracy of the impurity energy with respect to the total angular momentum of the polaron.  The angular momentum is partially shared 
by the excited bosons in a manner that is similar to the drag effect 
on the polaron momentum by a phonon cloud in the LLP theory 
for the electron-phonon system. 
\end{abstract}


\maketitle

\section{Introduction}
Polarons originally meant electrons dressed by locally excited phonons, 
which comprise one of the elementary excitations in ionic crystals. 
These excitations provide a widely applicable physics concept for quasiparticles 
in various environmental media \cite{Landau1,FPZ1,Mahan1,polaronreview1,LLP1}. 
Recently, many-body systems of trapped ultracold atoms allow us
to access the properties of such quasiparticles in a clean and controlled manner.  
Examples include studies on 
Bose-Einstein condensate (BEC) and Fermi polarons 
that are impurity atoms immersed in Bose-Einstein condensed atoms 
\cite{Cucchietti1,Sacha1,Tempere1,Casteels1,Rath1,Shashi1,Levinsen2,Dehkharghani5,
Ardila2,Yin1,Christensen1,Vlietinck1,Grusdt1,Grusdt3,Shchadilova1}
and degenerate Fermi atoms \cite{Chevy1,Schirotzek1,Schmidt1,Kohstall1,Koschorreck1,Vlietinck2,Massignan1,Yi1}, 
respectively, as well as on polarons in optical lattices \cite{Bruderer1}. 
Experimental realizations of BEC polarons were achieved first in a
weak coupling regime \cite{Catani1,Scelle1,Hohmann1}.   
Then, recent experiments in a strong coupling regime around the unitary limit 
have observed a behavior of the binding energy between an impurity 
and excited bosons in the BEC via radio frequency (RF) spectroscopy 
and in-situ imaging technology \cite{Jrgensen1,Hu1}.
These results show a smooth crossover from a weak mean-field to a strong molecular regime. 
In these experiments, a number of bosonic atoms are optically trapped to form a BEC, 
while an impurity atom immersed in the condensate starts to interact with bosons.  
Consequently, a polaron, i.e., an impurity accompanied by locally excited bosons, is formed.   The interaction between the impurity and bosons is characterized by an $s$-wave scattering 
length for low energy dynamics, while its sign and strength can be 
tuned by the Feshbach resonance from a weak- to a strong-coupling regime. 
The RF spectroscopy measures the energy shift between hyperfine states 
of the impurity due to the interaction, 
which corresponds to the binding (interaction) energy of the polaron. 

These experimental results for the BEC polarons are as a whole in agreement 
with theoretical predictions, which have been so far obtained entirely 
for spatially uniform systems. 
However, the real systems are in optical traps, 
which are well described by harmonic oscillator potentials.  
In the present study we investigate the properties of a BEC polaron in the case 
in which a single impurity atom and Bose condensed atoms are put in the respective 
axially symmetric harmonic potentials in three dimensions and interact 
attractively with each other in a regime where the coupling is weak or even intermediate, 
but is still far away from the unitarity.  
Since the axial component of the total angular momentum of the system is
conserved in such axially symmetric potentials, 
we focus on how the polaron's binding energy depends on a given total angular momentum 
of the impurity and excited bosons.  We also figure out detailed physics, 
such as a drag effect by excited bosons, that underlies the mean-field result 
in the trapped systems, although the mean-field result is occasionally referred
to as a reference theory at weak coupling.  

This paper is organized as follows: 
In Sec.~II we present the low energy effective Hamiltonian for a single 
impurity atom and bosons in axially symmetric harmonic potentials, 
and, by assuming that most of the bosons are in a BEC, i.e., at the 
lowest energy level, implement the Bogoliubov approximation to obtain a 
Yukawa-type interaction between the impurity and excited bosons.
In Sec.~III, for the state of the immersed impurity specified 
in terms of the harmonic oscillator eigenstates, 
we employ a variational method, {\it a la} Lee-Low-Pines (LLP) \cite{LLP1},  
to obtain the ground state of a polaron under fixed total angular momentum 
around the symmetric axis.  
In Sec.~IV, we present numerical results for the properties of polarons in 
various states by utilizing the parameter values that are used in 
experiments.  The last section is devoted to summary and outlooks.

\section{Effective Hamiltonian}
We consider a zero-temperature atomic many-body system of bosons and 
an impurity  that are trapped in axially symmetric harmonic potentials in three dimensions.   First, bosonic atoms  (denoted bya symbol `$b$') 
are condensed in the lowest energy level, 
and then a single impurity atom is introduced to interact attractively with bosons.  
Here we assume that the impurity is a fermion  (denoted by a symbol `$f$') for later convenience.  
A low-energy effective Hamiltonian for such a system is given by 
\beq
\mathcal{H}_{eff}\lk{\bf r}\rk
&=& 
H_{ho}^f\lk{\bf r}\rk
+\int_{{\bf r}'}\phi^\dagger({{\bf r}'})
\ldk
H_{ho}^b\lk{\bf r}'\rk
+g\delta^{(3)}({\bf r}-{\bf r}')
\rdk 
\phi({{\bf r}'})
\\
&=& 
H_{ho}^f\lk{\bf r}\rk
+\sum_{s} E_s^b  b_s^\dagger b_{s}
+g \sum_{s,s'} \phi_s^b\lk{\bf r}\rk^* \phi_{s'}^b\lk{\bf r}\rk b_s^\dagger b_{s'}, 
\label{Hamil1}
\\
H_{ho}^{b,f}\lk{\bf r}\rk
&=&
\frac{1}{2m_{b,f}}
\lk
p_z^2+p_r^2+\frac{L_z^2}{r^2}
\rk
+\frac{m_{b,f} \omega_{bt,ft}^2}{2} r^2 +\frac{m_{b,f} \omega_{b,f}^2}{2} z^2,
\eeq
where 
$p_z^2 \equiv -\frac{\partial^2}{\partial z^2}, \,  
p_r^2\equiv-\frac{1}{r}\frac{\partial}{\partial r}r\frac{\partial}{\partial r}, \, 
L_z\equiv-i\frac{\partial}{\partial \varphi}$  
is the angular momentum operator around the $z$ axis 
(the axial symmetry holds around the $z$ axis),  
${\bf r}=\lk r, \varphi, z\rk$ is  the cylindrical coordinate of the impurity,  
$\omega_{b}\lk\omega_{f}\rk$ and $\omega_{bt}\lk\omega_{ft} \rk$ are 
the frequencies of the harmonic potentials in $z$ and radial $r$ directions, respectively, 
for bosons (impurity), and $g=\frac{2\pi a_{bf}}{m_r}$ 
is the coupling constant between  the boson and the impurity  given 
in terms of an $s$-wave scattering length $a_{bf}$,  which is assumed to 
be negative and short, and the reduced mass 
$m_r=\frac{m_bm_f}{m_b+m_f}$ with $m_b$ ($m_f$) being the mass of boson (impurity) 
\cite{PethickSmith1}.  
We have ignored a possible boson-boson interaction, 
which would not bring qualitative changes in the present study 
as long as it is repulsive and so weak that the interaction energy is smaller than the trap frequency. 
In the case of relatively strong boson-boson interactions, 
i.e., $N_0 a_{bb}/\bar{a}\gg1$ 
where $N_0$ is the number of condensed bosons, 
$a_{bb}$ the boson-boson scattering length, 
and $\bar{a}$ the averaged harmonic amplitude \cite{PethickSmith1}, 
the condensate of trapped bosons is well described by semi-classical approximations such as the Thomas-Fermi approximation, 
while low energy excitations upon it become collective modes that still have discrete quantum numbers associated with symmetries of the system \cite{Dalfovo1,Pitaevskii1}.  
In contrast, the boson sector in our system with $a_{bb}=0$ is, 
for any state, in the quantum regime.  No semi-classical approximation 
is thus relevant, which allows us to easily examine how excited bosons 
with definite quantum numbers distribute around the impurity as will be seen later. 
As for the relevance to possible experiments, 
the present system is not necessarily academic,  
but vanishing $a_{bb}$ can be realized experimentally, e.g., for rubidium 
isotopes $^{85}$Rb and $^{87}$Rb, by using the Feshbach resonance 
\cite{abb1,abb2}, while the boson-impurity scattering length is left finite.  
It is also noted that the Bose collapse, 
not desired in this study, is prevented by the zero point energy in trap potentials 
even for negative $a_{bb}$ as long as it is sufficiently small  \cite{Mueller1}.

We have used the second quantized representation only for bosons, 
and expanded the boson field operator $\phi\lk {\bf r}\rk$  in terms
of the harmonic potential eigenfunctions $\phi_s^b\lk {\bf r}\rk$: 
\beq
\phi\lk {\bf r}\rk &=& \sum_s \phi_s^b\lk {\bf r}\rk b_s, 
\eeq
where $s$ denotes the quantum  number of the eigenstate 
whose single particle energy is given by $E_s^b$, and $b_s$ ($b_s^\dagger$) is 
the corresponding annihilation (creation) operator.  The explicit representation 
of  a set of  the quantum numbers  will be given  just below. 
We employ the abbreviation 
$\int_{\bf r}\equiv \int_{-\infty}^\infty {\rm  d} z 
\int_0^\infty {\rm  d} r r \int_0^{2\pi}{\rm d}\varphi$ and 
the unit in which $\hbar=1$ throughout the paper. 

\subsection{Bogoliubov-type approximation}
Since most of the bosons are in a BEC in the case of weak coupling and zero temperature, 
we implement the Bogoliubov-type approximation for 
the effective Hamiltonian, i.e., only interaction processes  involving 
the condensed bosons are taken into account: 
\beq
\mathcal{H}_{eff}
\simeq 
\mathcal{H} &=&  
H_{ho}^f\lk{\bf r}\rk+E_0^bN_0+gN_0 |\phi_0^b({\bf r})|^2
\nn
&&+\sum_{s\neq 0} E_s^b  b_s^\dagger b_{s}
+g\sqrt{N_0}\sum_{s\neq 0} 
\ldk \phi_0^b({\bf r})^* \phi_{s}^b({\bf r}) b_{s}
+\phi_s^b({\bf r})^* \phi_{0}^b({\bf r}) b_s^\dagger
\rdk, 
\label{Hamil2}
\eeq
where $s=0$ denotes the lowest energy level  at which the bosons are condensed, 
and $b_0, b_0^\dagger \simeq \sqrt{N_0}$, with $N_0$ being the number of  the 
condensed bosons.  
 We then express the Hamiltonian explicitly as  
\beq
{\mathcal H}
&=& 
\frac{1}{2m_{f}}
\lk
p_z^2+p_r^2+\frac{L_z^2}{r^2}
\rk
+\frac{m_f \omega_{ft}^2}{2} r^2 +\frac{m_f \omega_f^2}{2} z^2 
+E_0^bN_0+gN_0 |\phi_{0,0,0}^b(r,\varphi,z)|^2
\nn
&&
+{\sum_{n,n_t,m}}' E_{n,n_t}^b  b_{n,n_t,m}^\dagger b_{n,n_t,m}
\nn
&&
+g\sqrt{N_0}{\sum_{n,n_t,m}}'
\ldk \Phi_{n,n_t,m}(r,z) e^{im\varphi}   b_{n,n_t,m}
+\Phi_{n,n_t,m}^*(r,z) e^{-im\varphi}   b_{n,n_t,m}^\dagger
\rdk, 
\label{Hamil3}
\eeq
where the eigenenergies and eigenfunctions with a normalization factor 
${\mathcal N}$ for free bosons are given by 
\beq
E_{n,n_t}^b&=& \omega_b \lk n+\frac{1}{2}\rk +\omega_{bt} \lk n_t +1\rk, 
\label{eeb1}
\\
\phi_{n,n_t,m}^b({\bf r})&=& {\mathcal N}
e^{im\varphi} \Psi_{n_t,|m|}^b(r) \Psi_n^b(z) 
\label{esb1}
\eeq
with  the eigenfunctions  in $z$ and radial directions \cite{Bethe1}, 
\beq
\Psi_n^b(z) &\equiv&e^{-\frac{m_b \omega_{b}}{2}z^2} H_n\lk \sqrt{m_b \omega_{b}}z\rk, 
\\
\Psi_{n_t,|m|} ^b(r)&\equiv& 
r^{|m|}e^{-\frac{m_b \omega_{bt}}{2}r^2}
\frac{\lk\frac{n_t-|m|}{2}\rk! \Gamma\lk |m|+1\rk}
{\Gamma\lk |m|+1+\frac{n_t-|m|}{2} \rk}
L_{\frac{n_t-|m|}{2}}^{\lk |m| \rk}\lk m_b \omega_{bt}r^2\rk,
\eeq
 given in terms of the Hermite and Laguerre polynomials, $H_n(x)$ and 
$L_n^{(k)}(x)$, respectively.  The primary quantum numbers  in $z$ and 
radial directions are given by  $n=0, 1, 2, \cdots$ and $n_t=0, 1, 2, \cdots$, 
respectively, and the energy level $E_{n,n_t}^b$ is degenerate for  the 
eigenvalues of $L_z$: $|m|=n_t,n_t-2,n_t-4, \cdots ,1\, \mbox{ or }\, 0$.
We have also defined 
\beq
\Phi_{n,n_t,m}(z,r) &\equiv &e^{-i m\varphi} \phi_{n,n_t,m}^b({\bf r})  \phi_{0,0,0}^b({\bf r})^*, 
\eeq
and its complex conjugate $\Phi_{n,n_t,m}(z,r)^*$. 
Note that in the Hamiltonian (\ref{Hamil3}) the symbol ${\sum_{n,n_t,m}}'$ denotes 
the summation  over the boson's eigenstates  except $n=n_t=0$ 
 which corresponds to the state of the BEC. 

The eigenenergy and the corresponding eigenfunction of  a bare (free) impurity 
in a state  of $\lk {\nb,\ntb,\mb}\rk$  determined by the harmonic potential 
are obtained by  replacing the boson's mass and frequencies in 
(\ref{eeb1}) and (\ref{esb1})  with those of  the impurity: 
\beq
E_{\nb,\ntb}^f&=& \omega_f \lk n+\frac{1}{2}\rk +\omega_{ft} \lk n_t +1\rk, 
\label{eef1}
\\
\phi_{\nb,\ntb,\mb}^f({\bf r})&=& {\mathcal N}
e^{i\mb \varphi} \Psi_{\ntb,|\mb|}^f(r) \Psi_{\nb}^f (z). 
\label{esf1}
\eeq

\section{ Variational method {\it a la} LLP} 
We are interested in the ground state and  low-lying excited states of 
a single impurity atom  immersed in the BEC background,  as dictated by
the Hamiltonian (\ref{Hamil3}).  In order to  construct a solution with the total angular momentum 
in the $z$ direction conserved, 
we first use a gauge transformation $S$,  i.e., 
cranking of all bosons around the $z$ axis 
by $\varphi$, the angle of the impurity position \cite{Inglis1,Thouless1}, 
\beq
S&=&\exp\lk -i \varphi \sum_{n,n_t,m} m\, b_{n,n_t,m}^\dagger b_{n,n_t,m}\rk, 
\label{gt1}
\eeq
which transforms the operators as follows: 
\beq
S^{-1}b_{n,n_t,m}S &=& e^{-im\varphi}b_{n,n_t,m}, \quad
S^{-1}b_{n,n_t,m}^\dagger S= e^{im\varphi}b_{n,n_t,m}^\dagger, 
\\
S^{-1}\lk -i\partial_\varphi \rk S &=& 
-i\partial_\varphi - \sum_{n,n_t,m} m\, b_{n,n_t,m}^\dagger b_{n,n_t,m}. 
\eeq
Thus, the transformed Hamiltonian reads 
\beq
H'&\equiv&
S^{-1}\mathcal{H}S
\nn
&=&
\frac{p_r^2 }{2 m_f}
+\frac{1}{2m_f r^2}
\lk -i\partial_\varphi -\sum_{n,n_t,m} m\, b_{n,n_t,m}^\dagger b_{n,n_t,m}\rk ^2
+\frac{m_f \omega_{ft}^2}{2} r^2 
\nn
&&
+\frac{p_z^2 }{2 m_f}+\frac{m_f \omega_f^2}{2} z^2 
+E_0^bN_0+gN_0 |\phi_0^b({\bf r})|^2
+{\sum_{n,n_t,m}}' E_{n,n_t}^b  b_{n,n_t,m}^\dagger b_{n,n_t,m}
\nn
&&
+g\sqrt{N_0}{\sum_{n,n_t,m}}'
\ldk \Phi_{n,n_t,m}(r,z) b_{n,n_t,m}
+\Phi_{n,n_t,m}^*(r,z)  b_{n,n_t,m}^\dagger
\rdk. 
\label{transH1}
\eeq
In the gauge transformed system, the total angular momentum of the system 
(or a polaron) is converted to that of the impurity: 
\beq
S^{-1}\lk  -i\partial_\varphi +\sum_{n,n_t,m} m\, b_{n,n_t,m}^\dagger b_{n,n_t,m}\rk S=
-i\partial_\varphi = L_z, 
\eeq
which is a conserved quantity: $\ldk L_z, H' \rdk=0$.   In this respect we can 
define the angular momentum operator of  the impurity as 
\beq
L_{{\rm imp}, z}\equiv S^{-1}\lk -i\partial_\varphi \rk S 
=-i\partial_\varphi -\sum_{n,n_t,m} m\, b_{n,n_t,m}^\dagger b_{n,n_t,m}. 
\eeq
This is a very convenient property  when we describe the system with a conserved 
total angular momentum of $L_z$.  The cost we have to pay is that an interaction among 
bosons newly appears in the transformed Hamiltonian $H'$. 
Here we should note that a
more general transformation than (\ref{gt1}) is presented in the 
literature  \cite{Lemeshko3} for the description of a rotating impurity, 
so-called angulon \cite{Lemeshko4}, which  is characterized by 
the transfer of the angular momentum between the impurity and the  environmental bosonic degrees of freedom and by structural deformations of 
the bosonic distribution around the impurity. 
Although the system of an angulon assumes an infinite background space 
in contrast to our case of trapped atoms, emphasis is commonly put 
on the conserved quantity of the system, i.e. the total angular momentum. 

Now we take the expectation value of $H'$ over an impurity state, 
which we approximate to be an eigenstate  determined by the harmonic 
potential for  a bare impurity: $\phi^f_u(\br)$ that has a set 
of the quantum numbers $u=({\bar{n},\bar{n}_t,\bar{m}})$ in (\ref{esf1}),  
\beq
H_u
&\equiv& 
\int_{\bf r} \phi_u^{f*}({\bf r}) H' \phi^f_u({\bf r})
\\
&=& 
E^f_{u}
+E_0^bN_0+gN_0 C_{0,0;u,u}
+{\sum_{n,n_t,m}}' E_{n,n_t}^b  b_{n,n_t,m}^\dagger b_{n,n_t,m}
\nn
&&
+\frac{1}{2m_f} \left\langle \frac{1}{r^2} \right\rangle_u \lk -2\bar{m} \hat{m}+\hat{m}^2\rk
\nn
&&
+g\sqrt{N_0}{\sum_{n,n_t,m}}'
\ldk \bar{C}_{n,n_t,m; u}b_{n,n_t,m}
+\bar{C}_{n,n_t,m; u}^* b_{n,n_t,m}^\dagger
\rdk, 
\eeq
where we have introduced the operator for 
the total angular momentum ($z$ component) of excited bosons, 
\beq
\hat{m}\equiv {\sum_{n,n_t,m}} m\, b_{n,n_t,m}^\dagger b_{n,n_t,m}, 
\eeq
and defined the following quantities: 
\beq
\left\langle \frac{1}{r^2} \right\rangle_u 
&\equiv&
\int_{\bf r}\frac{1}{r^2} |\phi^f_u({\bf r})|^2, 
\\
\bar{C}_{n,n_t,m; u} 
&\equiv& \int_{\bf r} \Phi_{n,n_t,m}(r,z)  |\phi^f_u({\bf r})|^2, 
\\
C_{0,0;u,u} 
&\equiv& \int_{\bf r}  |\phi^b_0({\bf r})|^2  |\phi^f_u({\bf r})|^2
=\bar{C}_{0,0,0; u} . 
\label{c00uu}
\eeq
Note that the $H_u$ gives an effective Hamiltonian for excited bosons around the impurity 
whose angular momentum $\bar{m}$ is equivalent to the total  angular momentum of 
the system (a polaron) and  that this impurity state is only an approximate 
solution  in weak  coupling, while becoming the exact one when the interaction 
is  turned off.  We can improve the solution, e.g., by overlapping different 
impurity states with the same $\bar{m}$, or solving the impurity state  in a 
self-consistent potential generated by the excited bosons.  

Next we take the expectation value of $H_u$  over a coherent state of the excited bosons \cite{LLP1}, 
which is given by  a unitary transformation of the boson's Fock vacuum 
$|0\rangle$, 
\beq
|\phi_{ch}\rangle
&=&
\exp{{\sum_{n,n_t,m}}'
\lk f_{n,n_t,m}b^\dagger_{n,n_t,m} - f_{n,n_t,m}^* b_{n,n_t,m} \rk }  |0\rangle, 
\label{ch1}
\eeq
where $f_{n,n_t,m}$ (or its complex conjugate  $f_{n,n_t,m}^*$) is a 
 variational parameter.  Its physical meaning is the probability amplitude 
of  an excited boson being in a state  of $(n,n_t,m)$: 
\beq
f_{n,n_t,m}=\langle \phi_{ch}|b_{n,n_t,m}|\phi_{ch}\rangle. 
\eeq
Then the expectation value of $H_u$, i.e., the energy of  a polaron with 
a core impurity in a state $u=\lk \bar{n},\bar{n}_t,\bar{m}\rk$, 
becomes 
\beq
E_u&\equiv& \langle\phi_{ch}|H_u|\phi_{ch}\rangle
\nn
&=&
E^f_{\bar{n},\bar{n}_t}
+E_0^bN_0+gN_0 C_{0,0,u,u}
+{\sum_{n,n_t, m}}'E_{n,n_t}^b  |f_{n,n_t,m}|^2 
\nn
&&
+\frac{1}{2m_f} \left\langle \frac{1}{r^2} \right\rangle_u 
\ldk 
-2\bar{m} {\sum_{n,n_t, m}}' m|f_{n,n_t,m}|^2+{\sum_{n,n_t, m}}' m^2|f_{n,n_t,m}|^2
+\lk{\sum_{n,n_t, m}}' m|f_{n,n_t,m}|^2\rk^2
\rdk
\nn
&&
+g\sqrt{N_0}{\sum_{n,n_t, m}}' 
\ldk \bar{C}_{n,n_t,m; u} f_{n,n_t,m}
+\bar{C}_{n,n_t,m; u}^* f_{n,n_t,m}^*
\rdk. 
\label{exp1}
\eeq
Then taking the variation for the saddle-point condition as  
\beq
\frac{\delta E_u}{\delta f_{n,n_t,m}^*}
&=&
E_{n,n_t}^b f_{n,n_t,m}
\nn
&&
+\frac{1}{2m_f} \left\langle \frac{1}{r^2} \right\rangle_u 
\ldk 
-2\bar{m} m +m^2
+2m \lk {\sum_{n,n_t, m}}' m|f_{n,n_t,m}|^2\rk 
\rdk  f_{n,n_t,m}
\nn
&&
+g\sqrt{N_0}\bar{C}_{n,n_t,m; u}^* =0, 
\label{varia}
\eeq
we obtain a variational solution for the boson probability amplitude: 
\beq
f_{n,n_t,m; u}
&=& -g\sqrt{N_0}\bar{C}_{n,n_t,m; u}^* 
\ldk E_{n,n_t}^b 
+\frac{m^2-2\lk 1-\eta\rk \bar{m} m}{2m_f} \left\langle \frac{1}{r^2} \right\rangle_u 
\rdk^{-1}, 
\label{f1}
\eeq
where we have assumed that the excited bosons by the impurity partially share 
the total angular momentum of  a polaron $\bar{m}$ with a ratio $0\le \eta \le 1$ : 
\beq
\eta\, \bar{m}=\sum_{n,n_t,m} m|f_{n,n_t,m; u}|^2. 
\label{eta1}
\eeq 
We call $\eta$ the drag parameter, since  the above mechanism is very similar to 
the drag effect  in uniform systems on  a conserved polaron's total 
momentum \cite{LLP1}.  We can determine the numerical value of the parameter $\eta$ 
by solving Eq.~(\ref{eta1}) with the solution (\ref{f1}). 
It should also be noticed that the variational solution (\ref{f1}) is now a function 
of $\bar{m}$, and the dependence on $\mb$ brings an `anisotropy' in the summation of $m$ 
in (\ref{eta1}) to make the right hand side finite. 

Finally,  by plugging the solution (\ref{f1}) back into (\ref{exp1}), 
we obtain the expression  for the energy of the polaron in the state of 
$u=(\bar{n},\bar{n}_t,\bar{m})$  as 
\beq
E_u
&=& 
E^f_{\bar{n},\bar{n}_t}
+E_0^bN_0+gN_0 C_{0,0; u,u}
-\frac{1}{2m_f} \left\langle \frac{1}{r^2} \right\rangle_u 
\lk\sum_{n,n_t,m}m|f_{n,n_t,m; u}|^2\rk^2
\nn
&&
+g\sqrt{N_0}{\sum_{n,n_t, m}}' 
\bar{C}_{n,n_t,m; u} f_{n,n_t,m; u}
\nn
&\equiv& 
E^f_{\bar{n},\bar{n}_t}
+E_0^bN_0+E_{{\rm mf},u}
+E_{{\rm int},u},
\eeq
where we have defined the mean-field energy and the interaction energy, respectively,  as 
\beq
E_{{\rm mf},u}&=&gN_0 C_{0,0; u,u}=gN_0 {\bar C}_{0,0,0; u}, 
\label{mf1}
\\
E_{{\rm int},u}&=&-\frac{\bar{m}^2\eta^2}{2m_f} \left\langle \frac{1}{r^2} \right\rangle_u 
-N_0 g^2{\sum_{n,n_t, m}}'
\frac{|\bar{C}_{n,n_t,m; u} |^2}
{E_{n,n_t}^b +\frac{m^2-2\lk 1-\eta\rk \bar{m} m}{2m_f} 
	\left\langle \frac{1}{r^2} \right\rangle_u} 
\nn
&=&
-\frac{|\mb| \eta^2\omega_{ft}}{2} 
-N_0 g^2{\sum_{n,n_t, m}}'
\frac{|\bar{C}_{n,n_t,m; u} |^2}
{E_{n,n_t}^b +\frac{m^2-2\lk 1-\eta\rk \bar{m} m}{2|\mb|} \omega_{ft} }. 
\label{int1}
\eeq
In the interaction energy (\ref{int1}), the first term comes  from
decrease in the rotation energy of  the impurity by  the drag
effect, while the second term looks like a second-order perturbation 
 result that arises from virtually excited bosons,  although  
the non-perturbative nature is involved via the parameter $\eta$ 
that is self-consistently determined  from the variational solution. 
 In fact, the denominator of the second term can be decomposed 
up to the minus sign as 
\beq
\ldk E_{\nb,\ntb}^f+\frac{\mb_{\rm imp}^2}{2m_f}\left\langle \frac{1}{r^2}\right\rangle_u \rdk
-
\ldk E_{n,n_t}^b+E_{\nb,\ntb}^f+\frac{\lk \mb_{\rm imp}-m\rk^2}{2m_f}\left\langle \frac{1}{r^2}\right\rangle_u \rdk, 
\label{deco1}
\eeq
where the first term corresponds to  the impurity's single particle 
energy $E_{\nb,\ntb}^f$,  which is independent of $\mb$,  plus the  
rotation energy with $\mb_{\rm imp}\equiv \lk 1-\eta \rk \mb$, the angular 
momentum of  the impurity, while the second term  corresponds 
to an intermediate state  in which an excited boson of  $(n,n_t, m)$
takes the angular momentum $m$  off the impurity. 

 We conclude this section by considering transition amplitudes 
by single boson emission.  As  will be shown  in the next section, 
the ground state of  a polaron is given by $u=(\nb, \ntb, \mb)=(0,0,0)$, 
 while the other states correspond to excitations. 
The transition rate to  a lower energy state by single boson emission is proportional to 
the matrix element squared in perturbative treatment:  Defining a  polaronic 
state in $u=(\nb, \ntb, \mb)$ by  
$
|\psi_u\lk \br \rk\rangle
=S \phi_u^f\lk \br \rk  |\phi_{ch}\rangle_u, 
$
with $|\phi_{ch}\rangle_u$ denoting  the boson's coherent state (\ref{ch1}) 
 that has the solution  (\ref{f1}) used for $f_{s;u}$, we obtain the 
matrix element between  the initial polaronic state $u$ and  the final 
polaronic state $u'=(\nb', \ntb', \mb')$  accompanied by a boson emitted 
to a state $s=(n, n_t, m)$ as 
\beq
A_{u \rightarrow (u',s) }
&=&
\int_{\br}\langle\psi_{u'}\lk \br \rk| b_{s}  H_{\rm int}\lk \br \rk  |\psi_u\lk \br \rk\rangle
=
\delta_{m+\mb',\mb}\, 
g\sqrt{N_0}\sum_{s'}F_{s',s; u',u}, 
\label{amp1}
\eeq
where 
\beq
F_{s',s; u',u}&=& \exp\ldk -\frac{1}{2}\sum_{s''} \lk f_{s''; u'}-f_{s''; u}\rk^2 \rdk
\frac{1}{2\pi} 
\int_{\bf r} 
\Psi_{\nb'}^f\lk z \rk^*  
\Psi_{\ntb',\mb'}^f\lk r \rk^* 
\Psi_{\nb}^f\lk z \rk 
 \Psi_{\ntb,\mb}^f\lk r \rk \nn
&&\qquad\qquad\qquad\qquad
\times
\ldk \Phi_{s'}(r,z)  f_{s;u} f_{s';u} 
+
\Phi_{s'}(r,z)^* \lk  f_{s;u} f_{s';u'}^* + \delta_{s,s'}\rk 
\rdk 
\nn
&=&
\delta_{s,s'}
\int_{-\infty}^{\infty} {\rm d}z \int_{0}^{\infty} {\rm d}r r \, 
\Psi_{\nb'}^f(z)^*  
\Psi_{\ntb',\mb'}^f(r)^* 
\Psi_{\nb}^f(z)  
\Psi_{\ntb,\mb}^f(r)
\Phi_{s}(r,z)^* 
+{\mathcal O}(g^2),
\eeq
 and $H_{\rm int}$ is the last term of the right side of Eq.\ (\ref{Hamil3}). 
There appears a selection rule for the angular momentum $\delta_{m+\mb',\mb}$ in (\ref{amp1}), 
and if we consider as well the energy conservation $E^f_u=E^b_s+E_{u'}^f$
to leading order  in the coupling constant, 
the transition is allowed only in  the special case  of 
$\omega_f=\omega_b$ and $\omega_{ft}=\omega_{bt}$.   
This does not immediately imply the stability of  the polaronic state $u$, 
since there exist other decay processes, e.g., three-body loss \cite{Spethmann1}, 
which cannot be treated directly in our Hamiltonian.

\section{Numerical results and discussion}
We present numerical results  for the properties of  a polaron 
in the ground and  low-lying excited states, employing  
the parameter values that are used in the experiment for the 
boson-fermion mixture of ${}^{87}$Rb bosons in a BEC and ${}^{40}$K 
 impurity fermions \cite{Hu1}: 
\beq
N_0&=&\frac{2}{2\pi}\times 10^5,  \, 
a_{bf} =-187 a_0 \lk 1-\frac{-3.04}{B-546.62}\rk, \, a_{bb}=100 a_0, 
\\
\frac{\omega_b}{2\pi}&=&183\, {\rm Hz}, \, \frac{\omega_{bt}}{2\pi}=37\, {\rm Hz}, \, 
\frac{\omega_f}{2\pi}=281\, {\rm Hz}, \, \frac{\omega_{ft}}{2\pi}=50\, {\rm Hz},
\eeq
where $a_0=5.29177\times 10^{-11}$m, the Bohr radius, 
and the scattering length $a_{bf}$ is tunable by an external field $B$ (unit G). 
 For normalizations, we also use  the boson's inverse length scale 
and zero point energy: $k_{\rm ref} \equiv \lk 900 a_0\rk^{-1}$ and 
$E_{\rm ref}/ 2\pi \hbar \equiv 25 \,  {\rm kHz}$. 
 Incidentally, we can estimate from $N_0$ an average density of the BEC 
 in the oval sphere of harmonic amplitudes  as 
$n_0=N_0\ldk \frac{4\pi}{3}\sqrt{\frac{2\hbar}{\omega_b m_b}}
\sqrt{\frac{2\hbar}{\omega_{bt} m_b}}^2 \rdk^{-1}
= 2.389 \times 10^{14}  {\rm cm}^{-3}$,  which is of the order of 
a peak density in the experiment. 
Note that although these numbers lead to $N_0 a_{bb}/\bar{a}\sim 10$, 
which indicates that the bosonic sector is semi-classical, we will use 
them for numerical purpose in this study, except that we set $a_{bb}=0$. 
Nevertheless, in the experimental setup $a_{bb}$ can be vanishingly small by the Feshbach resonance as mentioned earlier,  and  
$\bar{a}$ is also tunable by the trap frequencies, while $N_0$ should 
be sufficiently large for the Bogoliubov-type approximation to be valid. 

\subsection{Mean-field energy and interaction energy}
\begin{figure}[h]
	\begin{center}
		\begin{tabular}{cc}
		 \resizebox{82mm}{!}{\includegraphics{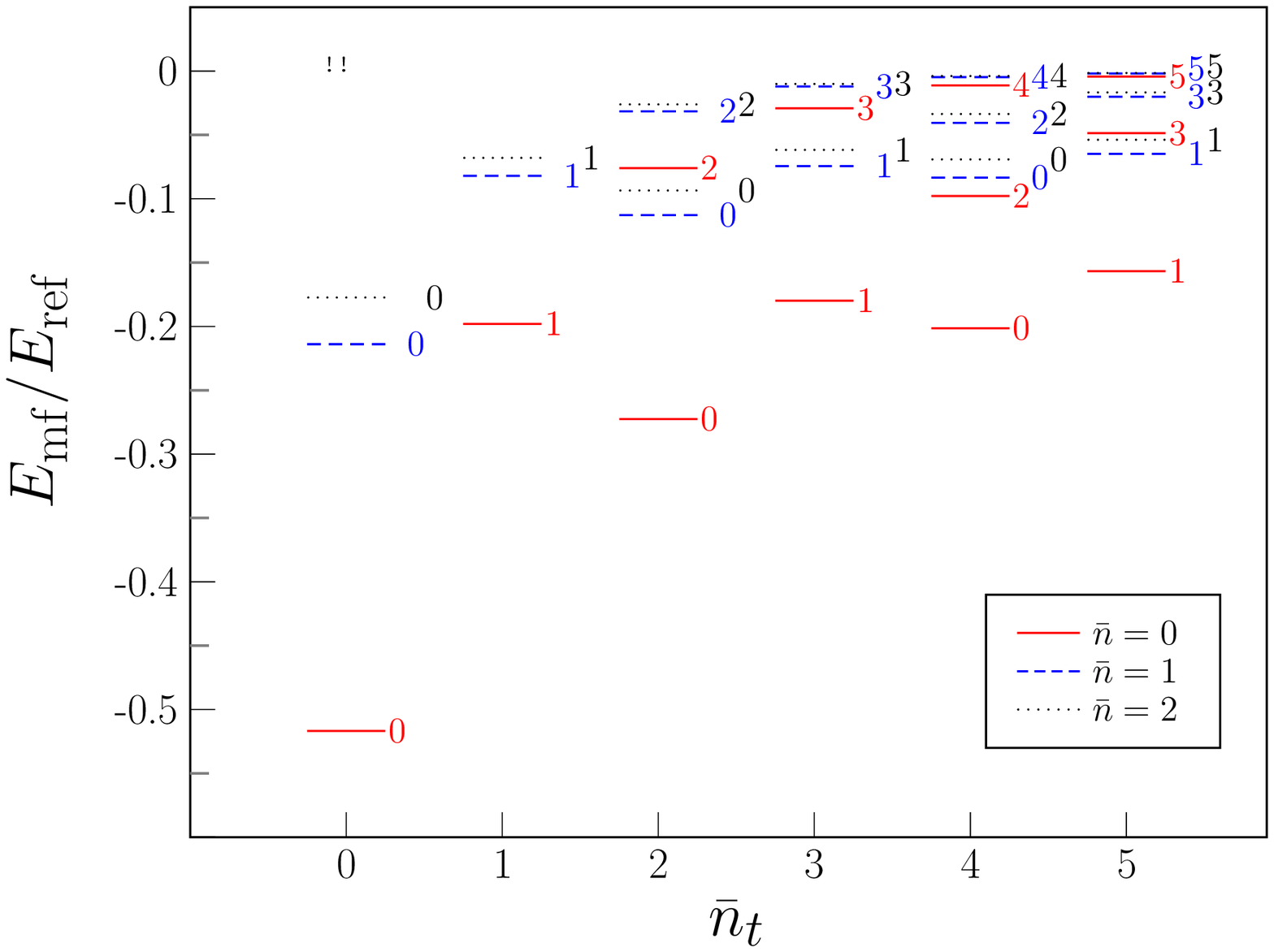}} \qquad \qquad   & 
		 \resizebox{82mm}{!}{\includegraphics{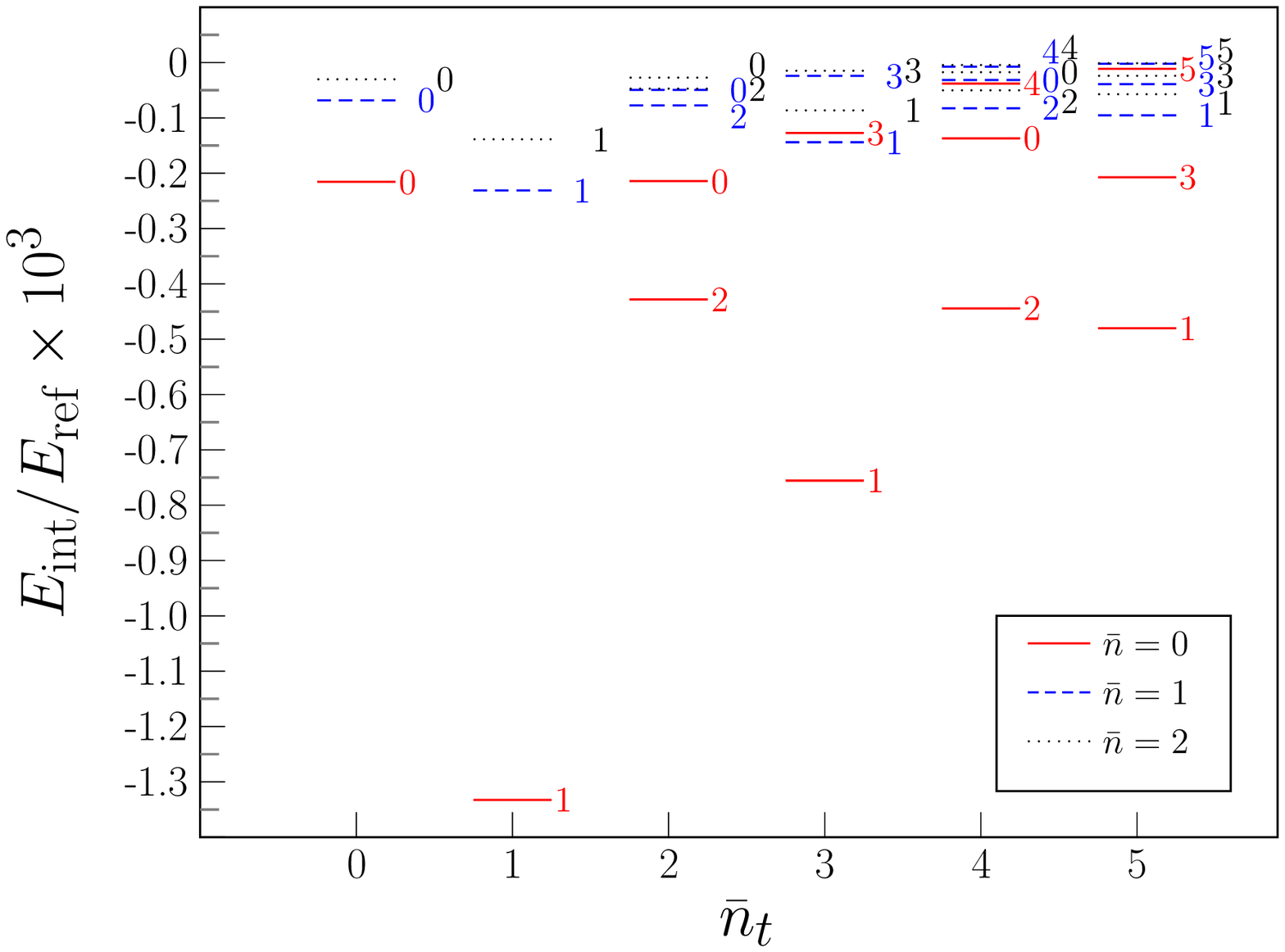}} 
		\end{tabular}
		\caption{(Color online)  Mean-field energy (\ref{mf1}) and  interaction energy (\ref{int1}), 
			plotted as a function of $\bar{n}_t$.  
            The number affixed to each bar denotes the value of $|\mb|$.}
		\label{fig1}
	\end{center}
\end{figure}

 We proceed to exhibit in Fig.~\ref{fig1} the mean-field and interaction 
energies, (\ref{mf1}) and (\ref{int1}), for  low-lying states of  the 
impurity: 
$\nb=0,1,2$, $\ntb=0,1,2,3,4,5$ ($|\mb|=\ntb, \ntb-2, \ntb-4, \cdots, 1\,  {\rm or}\, 0$),  
at $(a_{bf}k_{\rm ref})^{-1}=-900 a_0/187 a_0=-4.813$.  
 Here we have taken the sum  over $(n, n_t)$ 
up to $\lk n_{\rm max}=8, n_{t {\rm max}}=2n_{\rm max}\rk$ for boson excitations. 
We found a gradual convergence:  Increase in $n_{\rm max}$ by 50 ${\rm \% }$ 
results in a few ${\rm \% }$ changes in $\eta$ and $E_{\rm int}$. 
%
The mean-field energy dominates the binding energy of  a polaron in comparison with 
the interaction energy.  The ground state is given by the impurity state 
$(\nb=0,\ntb=0,\mb=0)$,  followed by the first and second excited states of 
$(\nb=0,\ntb=2,\mb=0)$ and $(\nb=1,\ntb=0,\mb=0)$, which can be accounted for 
by  a larger overlap  of the wave function of the BEC with those of 
$\mb=0$ impurity states than $\mb\neq 0$  (see the mean-field energy (\ref{mf1}) with (\ref{c00uu})). 
It should be noted that  the interaction energy for $\mb=0$  is equivalent 
to the second-order perturbation  due to virtual boson excitations, because 
$f_{n,n_t,m;u}=0$ for $m\neq 0$ (see Eq.\ (\ref{varia}) with
$\left\langle \frac{1}{r^2} \right\rangle_u = \frac{m_f \omega_{ft}}{|\bar{m}|}$) 
and hence only $m=0$ states for boson excitations  contribute to 
the summation in (\ref{int1}).  This,  in turn, leads to no drag effect $\eta=0$ 
for $\mb=0$  via (\ref{eta1}),  as will be seen again later. 

Looking into the dependence of the interaction energy on  the quantum 
numbers $u=(\nb,\ntb,\mb)$, we find that the states with $|\mb|=1$ for $\ntb=1, 3, 5$ 
gain relatively larger interaction energies than others.  This tendency 
can be understood from the fact that, in the summation  over a given 
$m$ and $-m$ in (\ref{int1}), the denominator gets smaller for larger values of 
$|\mb|$, while the overlap integral $\bar{C}_{n,n_t,m; \nb, \ntb, \mb}$ in the 
numerator gets smaller even more rapidly.  From the above observation,  
it turns out that the interaction between  the impurity and bosons breaks  the 
degeneracy of the single particle energy $E_u^f$ for a bare impurity state 
$u=(\nb, \ntb, \mb)$ with respect to $\mb$, and  that the mean-field and 
interaction energies split for $|\mb| \le \ntb$, where $\ntb \ge 2$ 
(the reflection symmetry still keeps  any $\mb$ and $-\mb$ states 
degenerate).  

Here we should mention that  
experimental observations of motional coherence of trapped impurity
atoms in the two lowest energy levels, in both the presence and 
the absence of the BEC background, has recently been achieved 
by motional Ramsey spectroscopy \cite{Scelle1},  which leads to 
the energy shift of the trapped impurity due to coupling
with the BEC background, i.e., 
``phononic Lamb shift'', for a weakly coupled impurity-boson interaction \cite{Rentrop1}.  
In analyzing this experimental result, Bogoliubov phonons of 
the BEC in free space and impurities trapped only in  
a single dimension were used.  In this analysis, 
the angular momentum is not relevant.  However, such experimental 
techniques could be utilized also for observations of 
fine level splittings between different angular momenta 
obtained in the present study.   

\subsection{Drag parameter $\eta$ and the number of excited bosons}
The above observation about  the $\mb\neq 0$ states also reflects
the drag parameter (\ref{eta1}) and the number of excited bosons due to  the 
interaction with the impurity, which is defined by 
\beq
N_{\rm exc}= \sum_{s\neq 0} \langle b_s^\dagger b_s\rangle={\sum_{n,n_t, m}}' |f_{n,n_t,m; u}|^2
\label{nexc1}
\eeq
with  $f_{n,n_t,m; u}$ given by (\ref{f1}).  
%
\begin{figure}[h]
	\begin{center}
		\begin{tabular}{cc}
			\resizebox{82mm}{!}{\includegraphics{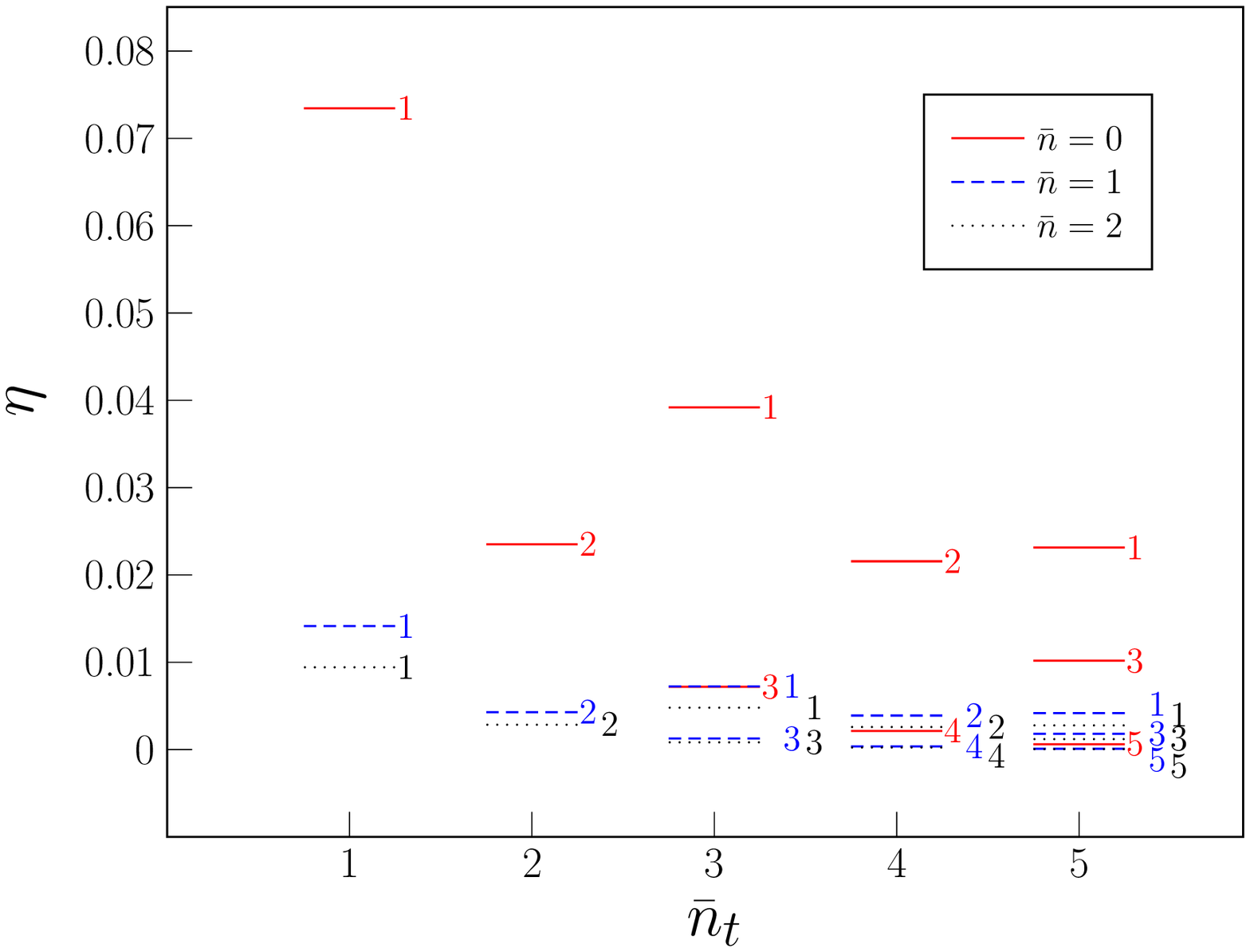}} \qquad \qquad  & 
			\resizebox{82mm}{!}{\includegraphics{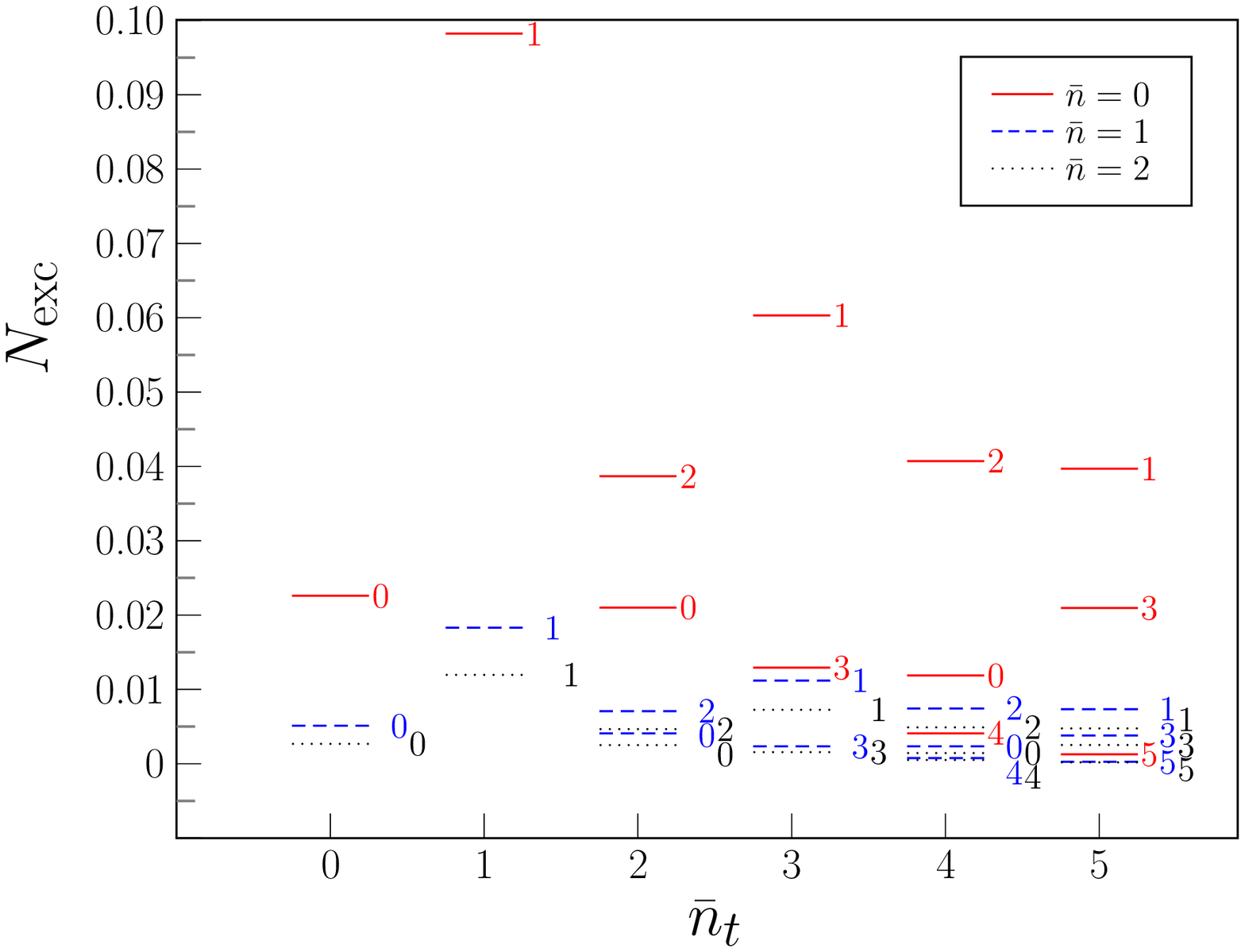}} 
		\end{tabular}
		\caption{(Color online)  Drag parameter (\ref{eta1}) and  the number of excited bosons  
                        (\ref{nexc1}), plotted as a function of $\bar{n}_t$.
                 The number affixed to each bar denotes the value of $|\mb|$.}
		\label{fig2}
	\end{center}
\end{figure}
As shown in Fig.~\ref{fig2}, the drag parameter $\eta$ becomes  nonzero 
for  the $\mb\neq 0$ states,  which takes on an especially 
 large value for  the $|\mb|=1$ states for a reason similar to 
the  case of the interaction energy.   Also, the distribution of 
$\eta$  has a pattern common to that of the number of excited bosons, 
 except that the latter is nonzero even for the $\mb=0$ states. 
This behavior is natural because $\eta$ is a fraction of the total angular 
momentum of  a polaron  carried by excited bosons. 

The number of excited bosons also tells us if the present Bogoliubov-type 
approximation is valid.  Figure \ref{fig2} shows that $N_{\rm exc}$ 
is about $0.1$ at most.  If $N_{\rm exc}$ is around unity, it obviously 
implies a breakdown of the present approximation, so that we need to restore 
the four-point residual interaction between  the impurity and excited 
bosons, which is responsible for correlation effects such as quasibound 
states among  the impurity and bosons. 

\subsection{Extension to strong coupling regime and lighter bosons}
\begin{figure}
	\begin{center}
		\begin{tabular}{cc}
			\resizebox{82mm}{!}{\includegraphics{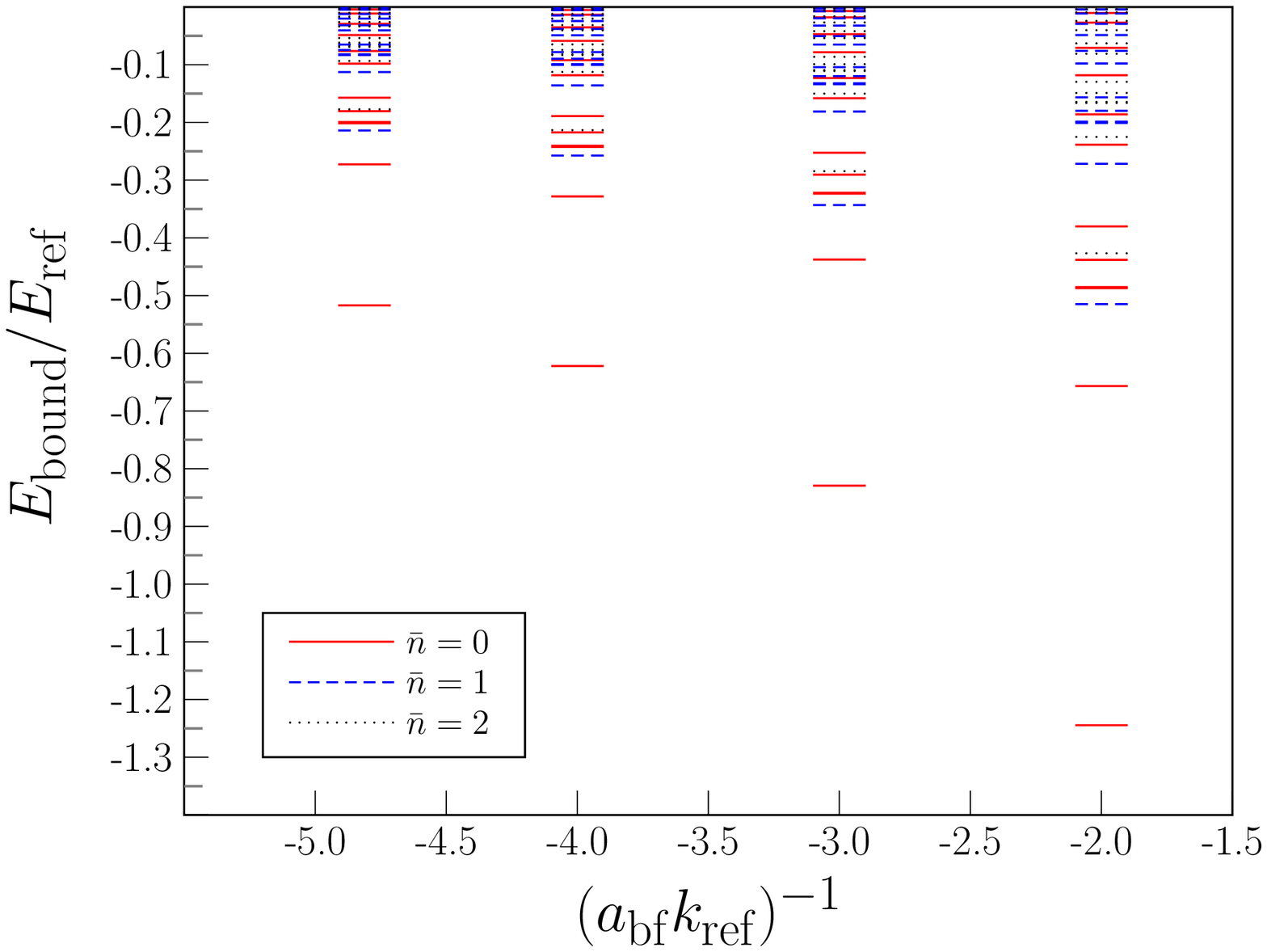}}  \qquad \qquad  & 
			\resizebox{82mm}{!}{\includegraphics{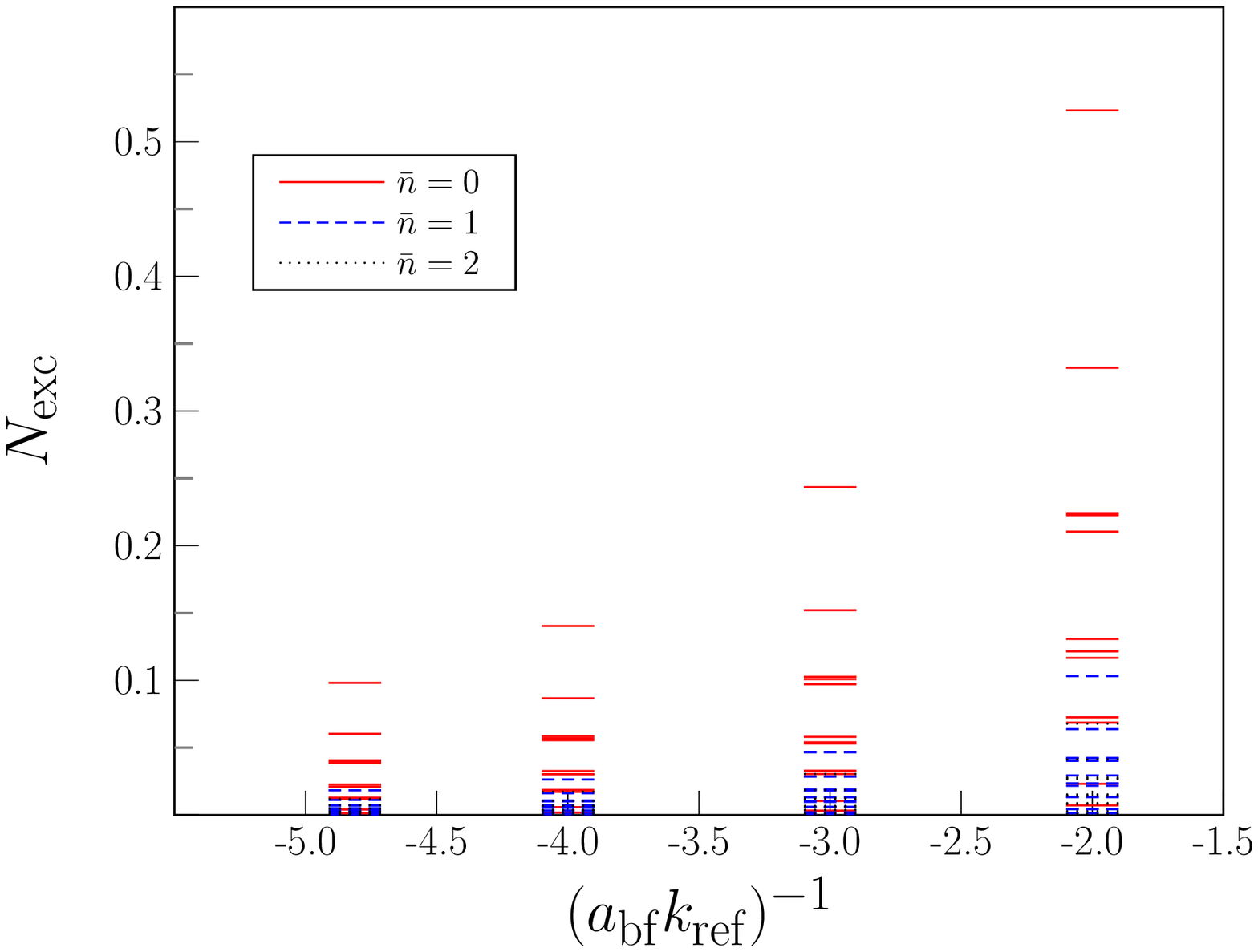}}
		\end{tabular}
		\caption{(Color online)  The polaron's binding energy (\ref{bou1}) 
     and  the number of excited bosons (\ref{nexc1}) for $\nb=0,1,2$, and $\ntb=0,1,2,3,4,5$, 
    plotted as a function of the inverse scattering length 
     $(a_{bf}k_{\rm ref})^{-1}=-4.813, -4, -3, -2$. 
     Indication of the quantum numbers $(\ntb,\mb)$ for each bar is omitted. }
		\label{fig3}
	\end{center}
\end{figure}

Hitherto we have fixed the coupling strength $g$ to a constant  that lies in a 
weak coupling regime.  Now we take some  typical values of  the coupling 
strength  to observe how the system transforms toward the strong coupling regime,
and  illustrate in Fig.~\ref{fig3} the coupling strength dependence of the 
binding energy of  a polaron, which is defined by 
\beq
E_{\rm bound} \equiv E_{\rm mf}+E_{\rm int}
\label{bou1},
\eeq  
and the number of excited bosons (\ref{nexc1}). 
The figure shows that the binding energy increases monotonically in magnitude
and  that there appears  a bunch of excited states above the ground state 
$(\nb,\ntb,\mb)=(0,0,0)$, which do not undergo level crossings  with increasing 
coupling strength. Moreover, the number of excited bosons, 
being largest for the state of $(\nb,\ntb,|\mb|)=(0,1,1)$, increases above $0.5$ 
at $(a_{bf}k_{\rm ref})^{-1}=-2$, which implies the limitation of the present approximation 
where the residual boson-fermion interaction is assumed to be negligible, i.e., $b_{s'}^\dagger b_s\ll 1$. 

It is also interesting to see how  our results are modified when bosons 
are  significantly lighter than  an impurity, 
since the LLP theory works better for relatively heavier impurities \cite{LLP1,Schweber1}. 
 \footnote{The LLP theory is originally applied to  a Fr\"{o}hlich-type 
 	Hamiltonian,  which is essentially  the same as ours (\ref{Hamil2}) 
 	in the absence of the residual four-point interaction, 
 	and gives the exact solution at heavy impurity limit.} 
 In the dimensionless expressions for the mean-field and interaction energies 
 (see (\ref{mf2}) and (\ref{int2}) in the Appendix), 
the  dependence on the boson and fermion mass can be factored out 
except for rescaling factors of  the boson's coordinates in  the 
wave functions, $z=R\zeta$ and $r=R_t\rho$,  which propagates to the 
overlap integrals  inherent in $\tilde{C}_{s,u}(R,R_t)$ in (\ref{int2}).
 In the case of  a heavy impurity  and/or light bosons, 
the overlap of the wave functions of  the condensed and excited bosons 
becomes larger at the origin in the integral, which  leads to a possible 
enhancement of these energies. 
%
\begin{figure}
	\begin{center}
		\begin{tabular}{cc}
			\resizebox{82mm}{!}{\includegraphics{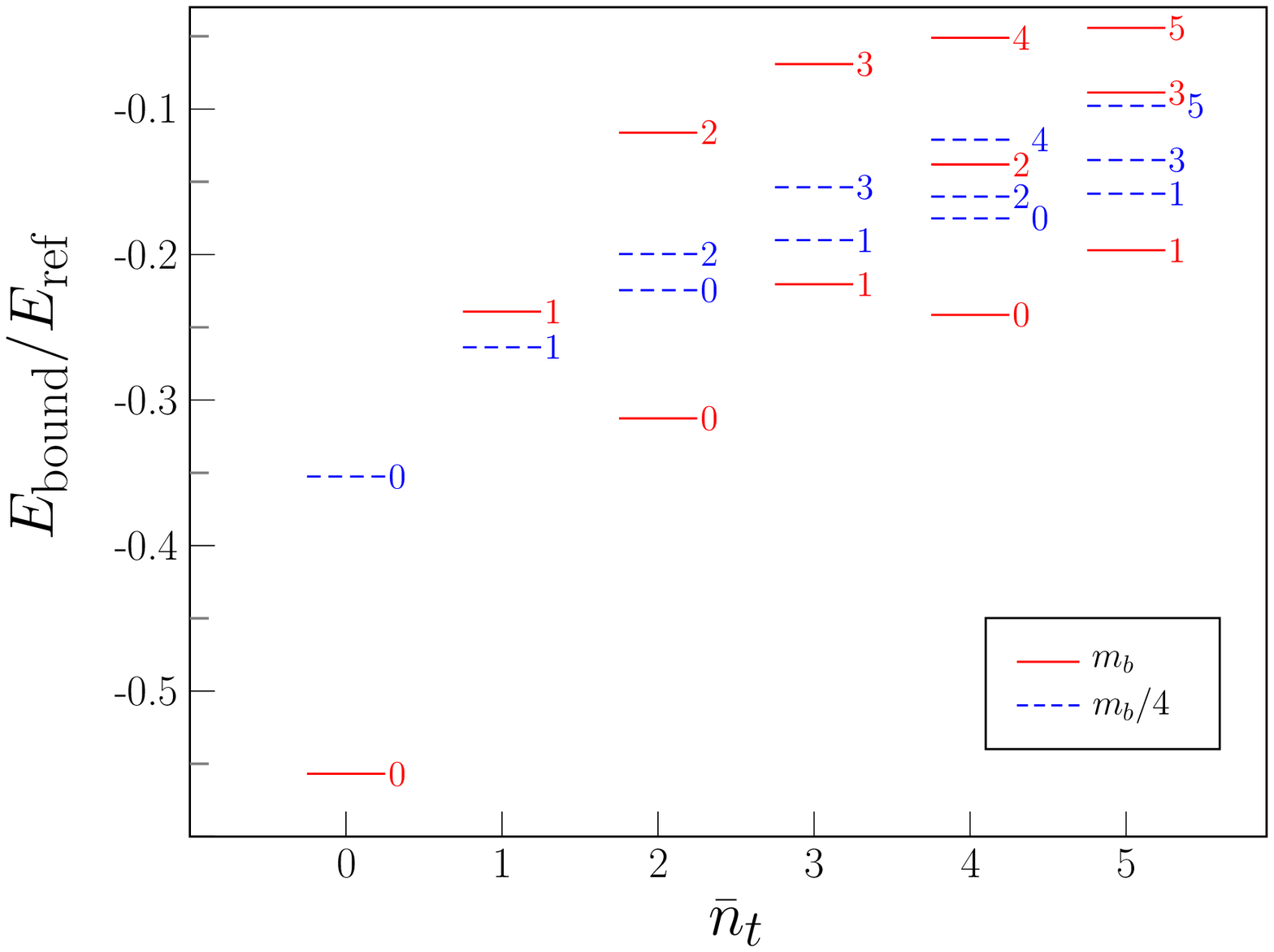}} \qquad \qquad & 
			\resizebox{82mm}{!}{\includegraphics{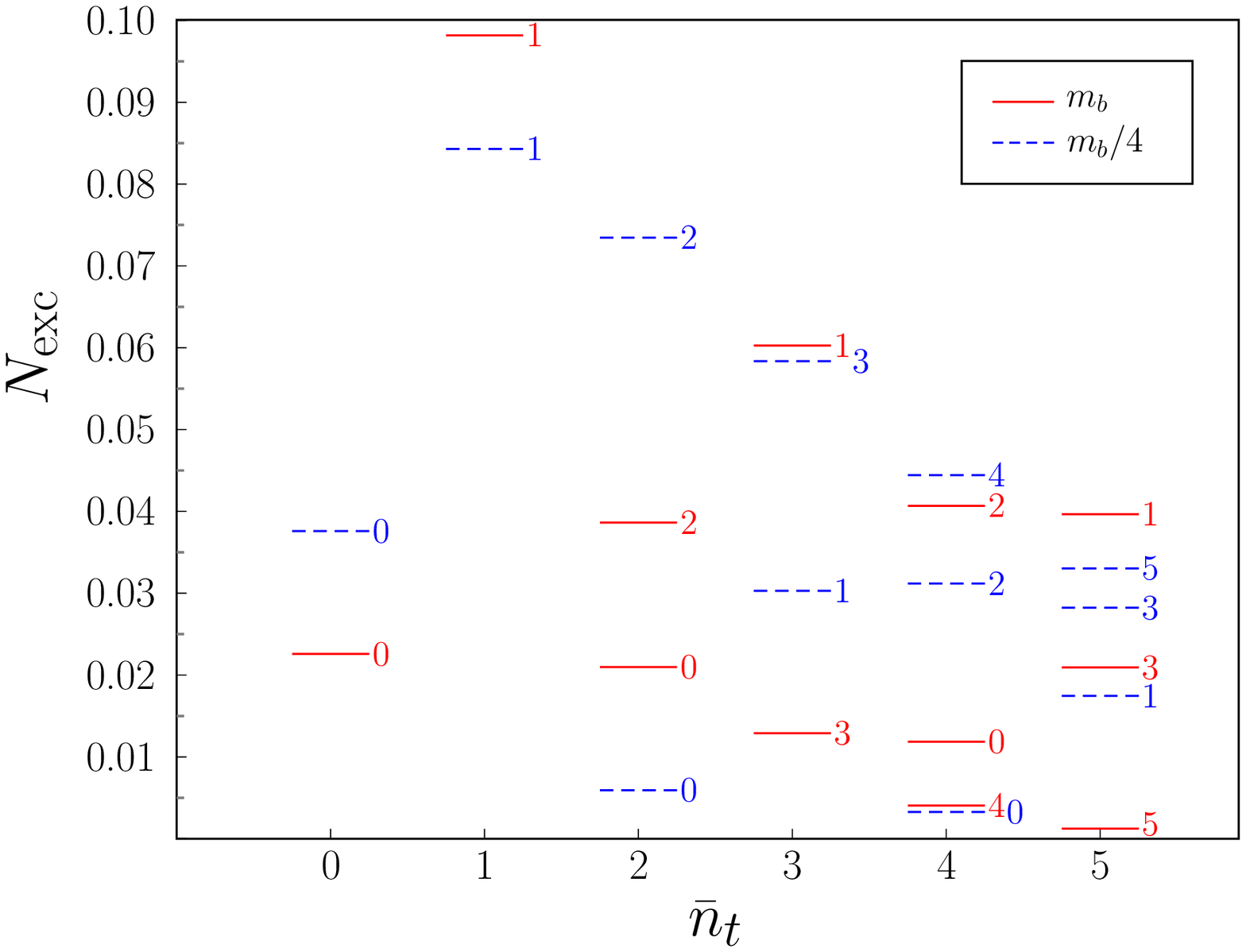}} 
		\end{tabular}
		\caption{(Color online)  The polaron's binding energy (\ref{bou1}) 
                       and the number of excited bosons (\ref{nexc1}) for
                         the boson mass of $m_b$ and $m_b/4$, 
                        $(a_{bf}k_{\rm ref})^{-1}=-4.813$, $\nb=0$, 
                  plotted as a function of $\bar{n}_t$.
              The number affixed to each bar denotes the value of $|\mb|$.}
		\label{fig4}
	\end{center}
\end{figure}
In Fig.~\ref{fig4} we show the binding energy  obtained by replacing
the boson mass $m_b$ with $m_b/4$, which is about a half of $m_f$.  
 We find from the figure that the virtual excitation of bosons is favored and 
 that a larger binding energy is gained in comparison with  the  
results  obtained for  the boson mass $m_b$ and the same coupling 
constant $1/a_{bf}k_{\rm ref}=-4.813$.  It is interesting to observe that the number of 
excited bosons gets larger for higher angular momentum states,  in 
contrast to the heavier boson case.   This is because the overlap 
integrals  associated with $\tilde{C}_{s,u}(R,R_t)$ in (\ref{prob2}) 
 are enhanced for lighter bosons.  These results  imply that 
 the salient features obtained in this study, such as the energy 
splitting and the drag effect for  nonzero angular momentum states, 
become more prominent  for $m_b\ll m_f$. 

\section{Summary and Outlooks}
We have investigated  the properties of a single BEC polaron trapped 
in an axially symmetric harmonic  potential in a weak to intermediate 
coupling regime, and for this purpose we have developed a formulation 
 based on an LLP-type  variational method \cite{LLP1}. 
We have obtained the mean-field energy  (\ref{mf1}) of ${\mathcal O}\lk g\rk$, 
whose magnitude is determined by the overlap of wave functions between  the 
impurity and BEC states, and dominates the total binding energy of  a polaron.  
We have also found that the interaction between  the impurity and excited bosons 
breaks the level degeneracy with respect to the total angular momentum  around 
the symmetric axis.  Our result shows that the interaction energy  (\ref{int1}) includes  an 
overall factor of ${\mathcal O}\lk g^2\rk$ and also a non-perturbative effect 
through the coefficient $\eta$,  the ratio of the angular momentum carried
by excited bosons to the total angular momentum $\mb$, but in the case of $\mb=0$ 
the interaction energy reduces to the second-order perturbation theory  
involving virtual boson excitations.  This situation is similar to the  LLP
description of spatially uniform electron-phonon systems,  according to 
which a drag parameter $\eta$ is introduced for the total momentum of  a polaron. 
Possible improvement beyond the present approximation to which we have 
simply taken a single eigenstate given by the harmonic potential 
for the impurity is that we can variationally determine the impurity state as well, 
especially for ${\bar m}=0$.

For  elaborate comparison with experimental results  in a weak to strong 
coupling regime, there remain several steps: When the $s$-wave scattering for 
boson-boson interactions is turned on, the excitation spectrum of  the boson 
sector would be modified, e.g., in the Thomas-Fermi regime the energy dependence 
on the quantum numbers  would be drastically changed \cite{Dalfovo1,Pitaevskii1}. 
For a microscopic description of such a semi-classical regime we need to 
solve, e.g., the Gross-Pitaevskii equation for the condensed and excited states, 
and obtain the effective low energy modes coupled to the impurity 
\cite{Japha1}. 
In the real experimental situations, 
impurity atoms themselves can form many-body systems 
and hence possible realizations of many-polaron systems if they are dilute 
enough for  the quasiparticle picture  to be valid.  In such cases we  
need to take into account the particle statistics of  impurities, e.g., Pauli 
blocking effect for fermionic impurities, in addition to finite temperature effects, 
and also  effects of interaction among polarons \cite{LDB2,Casteels5,LDB3,NY1,Naidon1},  
which may modify the polaron properties such as the spectral width and life time.  

For strong coupling  near the unitary limit of the boson-impurity interaction, 
or even for intermediate coupling, the present approximation  seemingly breaks down. 
In these regimes we have to restore the four-point boson-impurity interaction, 
which has been discarded in the present Bogoliubov-type approximation, 
but is responsible for scattering processes  between the impurity and excited 
bosons and,  around the unitary limit, for quasi bound states between them \cite{Rath1}. 
For a smooth description of trapped systems
from  a weak to strong  coupling regime,  it might be convenient to build 
such few-body correlations (quasi bound states) among  the impurity and bosons 
 into the present LLP-type approximation \cite{Shchadilova2}. 

Acknowledgments

E.N. is partially supported by  Grants-in-Aid for Scientific Research through Grant No.\ 16K05349 
provided by JSPS, and K.I. is partially supported by Grants-in-Aid for Scientific Research on
Innovative Areas through Grant No.\ 24105008 provided by MEXT.


\appendix
\section{Dimensionless variables for numerical calculations}
We introduce  the dimensionless variables, 
$\zeta \equiv \sqrt{m_f\omega_f}z$ and $\rho \equiv \sqrt{m_f\omega_{ft}}r$,  
and  the normalized eigenfunctions as 
\beq
\phi_{n,n_t,m}(\zeta,\rho,\varphi)&=&
\frac{1}{\sqrt{2\pi}}e^{im\varphi} \psi_n(\zeta) \psi_{n_t,|m|} (\rho), 
\\
\int_{-\infty}^{\infty}{\rm d}\zeta \psi_n^*(\zeta) \psi_{n'}(\zeta)&=&\delta_{n,n'}, 
\quad 
\int_{0}^{\infty}{\rm d}\rho \rho \psi_{n_t,|m|}^* (\rho) \psi_{n_t',|m|} (\rho)=\delta_{n_t,n_t'}, 
\eeq
where 
\beq
\psi_n(\zeta) &:=&
\sqrt{\frac{1}{2^n n! \sqrt{\pi}}} e^{-\frac{1}{2}\zeta^2} H_n\lk \zeta \rk, 
\\
\psi_{n_t,|m|} (\rho)&:=&
\sqrt{2\frac{\lk \frac{n_t-|m|}{2}\rk!}{\lk \frac{n_t+|m|}{2}\rk!}}
\rho^{|m|}e^{-\frac{1}{2}\rho^2}
L_{\frac{n_t-|m|}{2}}^{|m|}\lk\rho^2\rk. 
\eeq
Relations to  the normalized eigenfunctions 
with the original coordinates are given by 
\beq
\Psi_n(z) &=& \lk m_f\omega_f\rk^{1/4}\psi_n(\sqrt{m_f\omega_f}z), 
\\
\Psi_{n_t,m}(r)&=& \lk m_f\omega_{ft}\rk^{1/2} \psi_{n_t,m}(\sqrt{m_f\omega_{ft}}r). 
\eeq

For numerical calculations,  various quantities are given in terms of 
 the corresponding dimensionless quantities: 
\beq
\left\langle \frac{1}{r^2} \right\rangle_{\bar{n},\bar{n}_t,\bar{m}} 
&=&
m_f\omega_{ft}\int_0^\infty {\rm d}\rho \rho^{-1} |\psi_{\bar{n}_t,\bar{m}}(\rho)|^2
=\frac{m_f\omega_{ft}}{|\bar{m}|}, 
\eeq
where  $|\mb|= \ntb, \ntb-2, \ntb-4, \cdots, 1 \mbox{ or } 0$,
\beq
\bar{C}_{n,n_t,m; \bar{n},\bar{n}_t,\bar{m}} 
&\equiv& 
\int_{\bf x} 
\Phi_{n,n_t,m}({\bf x}) \left| \phi^f_{\bar{n},\bar{n}_t,\bar{m}}({\bf x})\right|^2
=
\sqrt{\omega_b m_b}\, \omega_{bt} m_b\, 
\tilde{C}_{n,n_t,m; \bar{n},\bar{n}_t,\bar{m}} \lk R,R_t\rk,  
\eeq
where 
$R=\sqrt{\frac{\omega_b m_b}{\omega_f m_f}}$, 
$R_t=\sqrt{\frac{\omega_{bt} m_b}{\omega_{ft}m_f}}$, 
and 
\beq
\tilde{C}_{n,n_t,m; \bar{n},\bar{n}_t,\bar{m}} \lk R,R_t\rk 
&\equiv&
\frac{1}{2\pi} 
\int_{-\infty}^{\infty}{\rm d}\zeta 
\psi_{0}^{*}(R\zeta)\psi_{n}(R\zeta) \left| \psi_{\bar{n}}(\zeta)\right|^2
\nn
&&
\times 
\int_{0}^{\infty}{\rm d}\rho \rho 
\psi_{0,0}^{*}(R_t\rho)\psi_{n_t,m}(R_t\rho) \left| \psi_{\bar{n}_t,\bar{m}}(\rho)\right|^2. 
\label{oi2}
\eeq

\subsection{Probability amplitude}
In the formula for the number of excited bosons (\ref{nexc1}), 
the probability amplitude squared can be rewritten 
in terms of the dimensionless variables as 
\beq
|f_{n,n_t,m; u}|^2
&=&
 N_0 G^2
\frac{|\tilde{C}_{n,n_t,m; \nb,\ntb,\mb}\lk R,R_t \rk|^2}
{\ldk \frac{\omega_b}{\omega_{ft}} \lk n+\frac{1}{2}\rk +\frac{\omega_{bt}}{\omega_{ft}} \lk n_t +1\rk+\frac{m^2-2\lk 1-\eta\rk \bar{m} m}{2|\mb|} \rdk^2}, 
\label{prob2}
\\
G&\equiv&\lk 2\pi a_{bf}\sqrt{\omega_bm_b}\rk \lk \frac{m_f}{m_b}+1 \rk R_t^2. 
\eeq
The above expression can also be used 
in the self-consistent Eq.~(\ref{eta1}) for $\eta$. 
Note that we cannot expand the right hand side of the self-consistent equation 
to the linear-order in $\mb$, unlike the drag parameter for the polaron's 
total momentum in the LLP theory for uniform systems. 

\subsection{Mean-field and interaction energies}
The mean-field and interaction energies for the state of $u=(\nb,\ntb,\mb)$ 
become 
\beq
\frac{E_{{\rm mf},u}}{\omega_{ft}}
&=& 
N_0 G\, 
\tilde{C}_{0,0,0; \bar{n},\bar{n}_t,\bar{m}} \lk R,R_t\rk, 
\label{mf2}
\\
\frac{E_{{\rm int},u}}{\omega_{ft}}
&=&
-\frac{|\mb|\eta^2}{2} 
-N_0 G^2 \sum_{n,n_t(\neq0),m} 
\frac{|\tilde{C}_{n,n_t,m; \nb,\ntb,\mb}\lk R,R_t \rk|^2}
{\frac{\omega_b}{\omega_{ft}} \lk n+\frac{1}{2}\rk +\frac{\omega_{bt}}{\omega_{ft}} \lk n_t +1\rk+\frac{m^2-2\lk 1-\eta\rk \bar{m} m}{2|\mb|} }. 
\label{int2}
\eeq

\end{document}